\documentstyle[12pt]{article}

\def\be{\begin{equation}}
\def\ee{\end{equation}}
\def\ben{$$}
\def\een{$$}
\def\ba{\begin{array}{c}}
\def\ea{\end{array}}

\begin{document}

\titlepage
\vspace*{2cm}

\begin{center}{\Large \bf
The Coulomb -- harmonic oscillator correspondence in ${\cal PT}$
symmetric quantum mechanics }\end{center}

\vspace{10mm}

\begin{center}
Miloslav Znojil
\vspace{3mm}

Nuclear Physics Institute
of Academy of Sciences
of the Czech Republic, 250 68 \v{R}e\v{z},
Czech Republic

\vspace{3mm}

and

\vspace{3mm}
G\'{e}za L\'{e}vai

\vspace{3mm}

Institute of Nuclear Research
of the Hungarian Academy of Sciences,
PO Box 51,
H-4001 Debrecen, Hungary

\end{center}

\vspace{5mm}

\section*{Abstract}

We show that and how the Coulomb potential $V(x)= Z\,e^2/x$ can
be regularized and solved exactly at the imaginary coupling
$Z\,e^2$.  The new spectrum of energies is real and bounded as
expected, but its explicit form proves totally different from
the usual real-coupling case.

\newpage

\section{Introduction}

\noindent Quantum mechanics often works with the exactly solvable
simplified models. For the precise fits of data or for some more
subtle quantitative analyses, unfortunately, the number of
solvable models is too limited. In $D$ dimensions, the only really
useful and easily tractable interactions are harmonic oscillators
and/or the central Coulomb well $V^{(Z)}(\vec{r}) =
-Z\,e^2/|\vec{r}|$. A new way out of this deadlock emerges within
the framework of the alternative, so called ``${\cal PT}$
symmetric" quantum mechanics. With its {\em complex} Hamiltonians
$H$ breaking both the parity ${\cal P}$ and the time-reflection
symmetry ${\cal T}$ and commuting only with their product ${\cal
PT}$, this formalism was proposed by Bessis \cite{Bessis} and by
Bender et al \cite{BB,Bender} as a possible way towards weakening
of the standard requirements of Hermiticity.

Several new exactly solvable ${\cal PT}$ symmetric models have
been proposed recently \cite{Cannata}. This is a promising
development with possible applications ranging from field theories
\cite{Milton} to supersymmetric models \cite{Andrianov} and from
quasi-classical methods \cite{Bender} to perturbation theory
\cite{pert}.

Even the solvable
harmonic oscillator itself acquires a richer
spectrum after its consequent ${\cal PT}$ symmetric regularization
in $D$ dimensions
\cite{PTHO}. The detailed structure of spectrum of this prominent
example does not in fact offer any really serious surprise. A
manifest violation of the parity ${\cal P}$ is compensated by an
emergence of the so called quasi-parity $q=\pm 1$ tractable as a
signature of two equidistant subspectra. The new quantum number
$q$ degenerates back to the eigenvalue of parity after a return
to the standard Hermitean and one-dimensional oscillator.

No immediate surprise emerges also for the quartic anharmonic
oscillator
\cite{BG}.
The situation only becomes less clear after one moves towards the
asymptotically vanishing models. They exhibit several
counterintuitive properties and open new
mathematical challenges \cite{ego}.  In particular,
the popular Coulomb potential did not not even seem particularly
suitable for any immediate ${\cal PT}$ symmetric regularization
\cite{QES}. A psychological barrier has been created by the
numerical and semiclassical studies of the general power-law
forces $V(x) \sim - (ix)^{\delta}$. They may be well defined
everywhere near the harmonic exponents $\delta = 2$, $\delta = 6$
etc \cite{sqw}.  At the same time, the related analyses hinted
that it is apparently difficult to move beyond the Herbst's
singularity located at $\delta = 1$ \cite{Herbst}.

In what follows, we intend to employ a slightly different strategy
and try to study the Coulomb problem directly, via its well known
correspondence to the harmonic oscillator. This correspondence is
based on an elementary change of variables. Its background dates
back to the nineteenth century mathematics and, in particular, to
the work of Liouville \cite{Olver}. The Newton's monograph
\cite{Newton} cites also Fivel \cite{Fivel} as a newer source of
the idea. In the contemporary literature (cf., e.g., \cite{Zhora}
for further references) people usually speak about the
Kustaanheimo - Steifel (KS) transformation \cite{KS}.
In all the implementations of this idea
the parameters appearing in the Coulombic and oscillator
problems are interrelated
of course.
Details will be mentioned below.
Preliminarily, let us only
warn the reader that all
the KS-type
mappings can also change the
dimensions and angular momenta and that
the energies of one
problem are related to the coupling constants of the other one and vice
versa. Within the "normal" quantum
mechanics, all this has
already been thoroughly discussed elsewhere: In ref.
\cite{varia}
for $D=3$ and in ref. \cite{Papp} for the
continuous transformation between Coulomb problems and harmonic
oscillators in various dimensions.

\section{Liouvillean changes of variables}

 \noindent The change-of-variable approach to the Coulombic
bound-state problem enables us to start directly from the harmonic
oscillator potential $W(r) =r^2$ or, in the present less traditional
context, from
its ${\cal PT}$ symmetric radial Schr\"{o}dinger equation
 \be \left[-\,\frac{d^2}{dr^2} +  \frac{l(l+1)}{r^2} +
W(r)\right]\, \chi(r) = \varepsilon^2
 \,\chi(r)
 \label{SEor}
  \ee
of ref.
\cite{PTHO}, using the complex coordinate $r = x
- i\,{c}$ with real $ x \in (-\infty,\infty)$ and with, say,
positive ${c} > 0$. This means that the integration path has been
shifted down from the position where it would cross the strong
centrifugal singularity.
Such a regularization preserves the asymptotic decrease of the
normalizable solutions.  Only in the limit ${c} \to 0$
and beyond the trivial one-dimensional case one has to omit all
the so called irregular solutions \cite{PRA}
defined by their ``physically
unnacceptable"  $\chi(r) \sim r^{-l}$ behaviour near
the origin.

Within the framework of the general Liouville method the change
of variables mediates a transition to the different potential
$V({t})$. It is easy to show that once we forget about boundary
conditions one merely has to demand the existence of an
invertible function $r=r({t})$ and its few derivatives $r'({t}),
\, r''({t}), \ldots$. Then, the explicit correspondence between
the two bound state problems may be {\em explicitly} given by
the elementary formulae.  From our original eq.  (\ref{SEor})
(i.e., in our case, harmonic oscillator) one obtains the new
(i.e., in our case, Coulombic) Schr\"{o}dinger equation
\be
\left[-\,\frac{d^2}{d{t}^2} + \frac{L(L+1)}{{t}^2}+
V({t})\right]\, \Psi({t}) = E \,\Psi({t})\
 \label{SE}
  \ee
with the new wave functions
  \be
\Psi({t})={ \chi[r({t})] / \sqrt{r'({t})}} \label{trenky}
  \ee
and with the new interaction and the new energies \cite{Olver},
 \ben
 \frac{L(L+1)}{{t}^2}+V({t})-E=\left [
 r'({t})
 \right ]^2
 \left \{ \frac{l(l+1)}{r^2({t})}+
 W[r({t})]-\varepsilon^2
 \right \}
+
 \frac{3}{4}
 \left [
 {r''({t}) \over r'({t})}
 \right ]^2
 -
 \frac{1}{2}
 \left [
 {r'''({t}) \over r'({t})}
 \right ].
  \een
Thus, it only remains for us to re-analyse the boundary
conditions.

\section{${\cal PT}$ symmetric KS transformation}

Without any
serious formal difficulties let us extend the scope of
the present considerations to all the singular forces
$\hat{W}(r) =W(r) + f/r^2$ and and/or $\hat{V}({t}) =V({t}) +
F/{t}^2$.  Both these central
forces may act in the respective $d$ and $D$ dimensions. This
means that
 \ben
 l(l+1) =
 \left \{ [j+(d-3)/2][j+(d-1)/2]+f \right \}
 \een
or rather
 \ben
 (l+1/2)^2 \ (= \alpha^2) =
 \left [ j+(d-2)/2]\right ]^2 + f
 \een
(with Re $\alpha > 0$) and
 \ben
 (L+1/2)^2 \ (= A^2)  =
 \left [ J+(D-2)/2]\right ]^2 + F
 \een
(with Re $A > 0$)
where the partial waves are numbered
by the respective integers
$j = 0, 1, \ldots$ and $J = 0, 1, \ldots$.

An important simplification of our effort is provided by our
knowledge of the complete harmonic oscillator solution as
derived in ref. \cite{PTHO}. Its two equidistant subsets of
energies
 \ben
\varepsilon^2=  \varepsilon_{(n,q)}^2 = 4n+2-2\,q\,\alpha,
\ \ \ \ \ \ \ \ q = \pm 1, \ \ \ \ n = 0, 1, \ldots \,
\een
correspond to the two families of the Laguerre-polynomial wave
functions
 \ben
 \chi_{(n,q)}(r) = {\cal N}\,r^{1/2-q\,\alpha}\,e^{-r^2/2}
\,L^{(-q\,\alpha)}_{n}(r^2)\, .
 \een
Integration path $r = x - i\,c$ lies in the lower half of the
complex plane and does not change after the subsequent ${\cal P}$
and ${\cal T}$ transformations $r=r(x) \to -r$ and $-r \to
-r^*=r(-x)$.

In the spirit of the above-mentioned KS mapping of harmonic
oscillators on Coulombic bound states we now have to define a
complex variable $t$ as a re-scaled square of
$r(x)$ such that the resulting path $t(x)$ remains ${\cal
PT}$ invariant. Our requirement implies that the complex
plane of $r$ will cover twice the complex plane of $t$. In such
an arrangement, our lower half plane of $r$ should cover the
Riemann sheet given as a whole plane of $t$ which is cut upwards
from the origin.
In the polar representation,
one has $r \sim \exp (-i\,\varphi)$ mapped upon
$t \sim \exp (-2\,i\,\varphi)$ with $\varphi \in (0, \,\pi)$.
Once we introduce a suitable free parameter $\kappa>0$ we can
put, say, $r^2 = 2\,\kappa^2 z$ and then rotate the $z-$plane
(which is cut, by construction, along the real and positive
semi-axis) by the angle $\pi/2$ giving $t = i\,z$.  Our final
recipe
\be
r^2=-2\,i\,\kappa^2t
\ee
maps the above-mentioned straight line $r(x) = x - i\,c$ upon a
curve $t(x) = u + i\,v$. Its real part $u=u(x)=x\,c/\kappa^2$
and imaginary part $v=v(x)= (x^2-c^2)/2\,\kappa^2$ form, as
required, a ${\cal PT}$ symmetric and upwards-oriented parabola
$v = -c^2/2\,\kappa^2+ (\kappa^2/2\,c^2)\,u^2$ in complex plane.
Obviously, a small asymptotic deformation of our original curve
$r(x)$ with the modified shifts $c=c(x) \sim 1/x^{1+\eta}$ would
transform the parabola to a pair of lines which are parallel to
the cut in the asymptotic domain of $|x| \gg 1$.

Having achieved a ${\cal PT}$ symmetry in the complex
plane of $t$, we may move to the (trivial) insertions and
conclude that all the above-mentioned harmonic oscillator
bound-state solutions are in a one-to-one correspondence to the
solutions of the Coulombic Schr\"{o}dinger equation (\ref{SE}),
\be
\left[-\,\frac{d^2}{d{t}^2} + \frac{L(L+1)}{{t}^2}
+i\,\frac{Z\,e^2}{t}
\right]\, \Psi({t}) = E \,\Psi({t})\ ,
\ \ \ \ \ \ \ \ \ t = u(x) + i\,v(x), \ \ \ \ \ x \in
I\!\!R.
 \label{SEc}
  \ee
The underlying assignment of constants is such that $\alpha =
2\,A$ while $\kappa$ itself becomes $n-$dependent,
$\kappa^2=2\,Z\,e^2/\varepsilon^2=Z\,e^2/(2n+1-2\,q\,A)$.  In
full detail one gets the new, Laguerre-polynomial wave functions
\be
 \Psi_{(n,q)}(t) = {\cal M}\,t^{1/2-q\,A}\,e^{i\,\kappa^2t}
\,L^{(-2\,q\,A)}_{n}(-2\,i\,\kappa^2t)\,
\ee
and their energy spectrum specified by the elementary formula
 \be
E_{(n,q)}=  \kappa_{(n,q)}^4 =
\frac{Z^2e^4}{
(2n+1-2\,q\,A)^2}
\ \ \ \ \ \ \ \ q = \pm 1, \ \ \ \ n = 0, 1, \ldots \, .
\label{energs}
\ee
This is our main result.

\section{Discussion}

\subsection{Consequences of the curvature of our integration path}

The latter two formulae exhibit several unusual features.  The
first concerns the asymptotics of the wave functions which are
determined by the decreasing exponential $\exp({i\,\kappa^2t})$.
Its form re-confirms the correctness of the above, slightly
counter-intuitive KS-dictated choice of our ${\cal PT}$
symmetric integration path. Asymptotically, it encircles more or
less closely the positive imaginary axis in $t$ plane.  This
clarifies the apparent paradox.

The second unexpected result is the positivity and unusual
$n-$dependence of the energies.  This can be related to the
choice of the KS integration path  $t(x)$ again. In the very
vicinity of the origin, one can visualize this path as a
circle with radius $\sigma$,
\ben
u^2(x) + v^2(x) = \sigma^2, \ \ \ \ \ \ |x| \ll 1.
\een
From the appropriate definitions we get the formula
\ben
\sigma = c^2(0)/2\,\kappa^2_{(n,q)} + {\cal O}(x^2)
\een
and see that this radius is $n-$dependent and increases with the growth of this principal quantum
number,  $\sigma \sim
n\,c^2(0)/Z\,e^2$, $n \gg 1$.
As a consequence, an ``effective charge" of
our ${\cal PT}$ symmetric Coulomb potential appears to decrease
with $n$ since
\ben
\left |\frac{i\,Z\,e^2}{t} \right | \sim \frac{Z\,e^2}{\sigma}
+{\cal O}(t)={\cal O}(1/n).
\een
This offers a ``rule-of-thumb" guide to the unusual and
certainly counterintuitive $n-$dependence of the energy levels
(\ref{energs}).  Of course, in practice, a preferred
integration path will be $n-$independent.  In such
a case, the $n-$dependence re-appears in the
small-$x$ deformation of the
initial harmonic-oscillator path with $c=c_n(x)={\cal O}(1/n)$.
Such a flexible transfer of
the excitation-dependence
throws also a new light on the complexified
KS transformation itself.

\subsection{``Flown-away" energies and unavoided crossings}

Let us return in more detail to the $A-$dependence of our
energies (\ref{energs}).
Firstly, we notice their power-law dependence on $n$ and $A$
(with exponent = -2)
as somewhat similar to
the spectra in ${\cal PT}$
 symmetric oscillator well
\cite{PTHO}
(with exponent = +1)
  and in the Morse potential
of ref. \cite{Morse}
(with exponent = +2).

In the present case, obviously, we have to
distinguish between the two separate families $E_{(n,q)}$
with $q=+1$ (cf.
Figure 1) and $q=-1$ (cf. Figure 2).  The latter set is,  up to
its sign, analogous to the ordinary Coulombic
spectrum.
By far not so the former one.  Its energies enrich and dominate
the spectrum.  The $n_{div}-$th
energy ``flies away" and
disappears from the spectrum at
$A_{div}=n_{div}+1/2$.
Moreover, at all the positive integers and
half-integers $A$ one encounters the {\em
unavoided} level crossings.
In contrast to the harmonic case, they appear
at {\em both} the opposite and
identical (viz., positive)
quasiparities $q$. The former case takes
place at $A=A_{crit}=(n-n')/2$ while in the latter case we must
fulfill the condition $A=A_{crit}=(n+n'+1)/2$.  A sample of this
phenomenon is given in Figure 3.

Formally the unavoided crossings generate certain identities
which connect different Laguerre polynomials (cf. their sample
in ref. \cite{PTHO}). In applications, these ``critical" cases are
not exceptional at all.  For
$F=0$ forces without a spike, the critical integer
or half-integer coordinates $A_{crit}=J-1+D/2$
correspond {\em precisely} to the physical (namely, integer)
dimensions $D$ and partial waves $J$.

In the conclusion, let us
not forget about many open questions.  {\it
Pars pro toto}, one could mention
a not yet clear possibility of
re-interpretation of our bound states, say, in the limit $t \to
real$, i.e., beyond the mathematical and apparently natural
boundaries of our present approach.
Moreover, we must keep in mind that
via our complexification of the coordinates we of course broke
their immediate connection to any standard
$D-$dimensional
problem.  In this sense, our $F=0$ and $F\neq 0$ Hamiltonians differ
just in an inessential way. One could even prefer
the latter, Kratzer-like option as a model which is formally
simpler, due to the generic
absence of the puzzling unavoided crossings.
At $F \neq 0$ the structure of the spectrum of our present
Coulomb model becomes also richer and, in this sense,
more interesting.

\section*{Acknowledgements}
M. Z. thanks for
the hospitality
of the
Institute of Nuclear Research
of the Hungarian Academy of Sciences,
in Debrecen, and appreciates
the support by the
grant Nr. A 1048004 of GA AS CR.
G. L. acknowledges the OTKA grant no. T031945.


\newpage

 \begin{thebibliography}{00}

\bibitem{Bessis}
D. Bessis, private communication (1992).

\bibitem{BB}
C. M. Bender and S. Boettcher, Phys. Rev. Lett. { 24} (1998) 5243;

\bibitem{Bender}
E. Delabaere and F. Pham, Phys. Lett. A 250 (1998) 25;

C. M. Bender, S. Boettcher and P. N. Meisinger, J. Math. Phys. 40
(1999) 2201.

\bibitem{Cannata}
F. Cannata, G. Junker and J. Trost, Phys. Lett. { A 246} (1998)
219;

B. Bagchi and R. Roychoudhury,  J. Phys. A: Math. Gen. 33 (2000)
L1;

M. Znojil,
 J. Phys. A: Math. Gen. 33 (2000)
L61.

\bibitem{Milton}
C. M. Bender and K. A. Milton, Phys. Rev. D 55 (1997) R3255
 and 57 (1998) 3595.

\bibitem{Andrianov}
A. A. Andrianov, M. V. Ioffe, F. Cannata and J. P. Dedonder, Int.
J. Mod. Phys. A 14 (1999) 2675

\bibitem{pert}
E. Caliceti, S. Graffi and M. Maioli, Commun. Math. Phys. 75
(1980) 51;

F. Fern\'andez, R. Guardiola,  J. Ros and M. Znojil,  J. Phys. A:
Math. Gen. 31 (1998) 10105;

C. M. Bender and G. V. Dunne, J. Math. Phys. 40 (1999) 4616;

E. Caliceti, arXiv: math-ph/9910001.

\bibitem{PTHO}
M. Znojil, Phys. Lett. A 259 (1999) 220.

\bibitem{BG}
V. Buslaev and V. Grecchi,
 J. Phys. A: Math. Gen. 26 (1993) 5541;

C. M. Bender and K. A. Milton, J. Phys. A: Math. Gen. 32 (1999)
L87;

M. Znojil, J. Phys. A: Math. Gen. 32 (1999) 7419.

\bibitem{ego}
M. Znojil,
arXiv:
quant-ph/9912027,
quant-ph/9912079 and
 math-ph/0002017.

\bibitem{QES}
M. Znojil, J. Phys. A: Math. Gen. 32 (1999) 4563.

\bibitem{sqw}
C. M. Bender, S. Boettcher, H.F. Jones and Van M. Savage, J. Phys.
A: Math. Gen. 32 (1999) 6771;

C. M. Bender,F. Cooper, P. N. Meisinger and V. M. Savage,
 Phys. Lett. { A 259} (1999)
224.

\bibitem{Herbst}
I. Herbst, Commun. Math. Phys. 64 (1979) 279.

\bibitem{Olver}
J. Liouville, J. Math. Pures Appl. 1 (1837) 16;

G. Jaff\'e, Z. Phys. 87 (1933) 535;

F. W. J. Olver, Introduction to Asymptotics and Special Functions
(Academic, New York,  1974), Chap. VI.


\bibitem{Newton}
R. G. Newton, Scattering Theory of Waves and Particles (Springer,
Berlin, 1982), p. 442, problems 17 and 18.


\bibitem{Fivel}
D. I. Fivel, Phys. Rev. 142 (1966) 1219.

\bibitem{Zhora}
E. G. Kalnins, W. Miller Jr. and G. S. Pogosyan,
arXiv:
quant-ph/9906055.

\bibitem{KS}
P. Kustaanheimo and E. Steifel,
J. Reine Angew. Math. 218 (1965) 204.

\bibitem{varia}
M. Znojil, J. Phys. A: Math. Gen. 27 (1994) 4945.

\bibitem{Papp}
G. L\'{e}vai, B. K\'{o}nya and Z. Papp, J. Math. Phys. 39 (1998)
5811.

\bibitem{PRA}
M. Znojil,
arXiv:
quant-ph/9811088, to appear in Phys. Rev. A.

\bibitem{Morse}
M. Znojil,
Phys. Lett. A. 264 (1999) 108.


\end{thebibliography}
\end{document}